\documentclass[9pt,conference]{IEEEtran}
\IEEEoverridecommandlockouts
\usepackage{cite}
\usepackage{amsmath,amssymb,amsfonts}
\usepackage{algorithmic}
\usepackage{graphicx}
\usepackage{textcomp}
\usepackage{xcolor}
\usepackage{caption, subcaption}

\def\BibTeX{{\rm B\kern-.05em{\sc i\kern-.025em b}\kern-.08em
    T\kern-.1667em\lower.7ex\hbox{E}\kern-.125emX}}
\begin{document}

\title{Perceptual Audio Coding: A 40-Year Historical Perspective}


\author{

\IEEEauthorblockN{Jürgen Herre}
\IEEEauthorblockA{\textit{International Audio Laboratories Erlangen$^1$}
\thanks{$^1$ A joint institution of the Friedrich-Alexander Universität Erlangen-Nürnberg (FAU) and Fraunhofer IIS.}\\
Erlangen, Germany \\
juergen.herre@audiolabs-erlangen.de}
\and
\IEEEauthorblockN{Schuyler Quackenbush}
\IEEEauthorblockA{\textit{Audio Research Labs} \\
Westfield, NJ \\
srq@audioresearchlabs.com}
\and
\IEEEauthorblockN{Minje Kim$^2$}\thanks{$^2$ Supported in part by ETRI grant funded by the Korean government (24ZC1100; ``The research of the basic media contents technologies.")}
\IEEEauthorblockA{\textit{University of Illinois at Urbana-Champaign} \\
Urbana, IL \\
minje@illinois.edu}
\and
\IEEEauthorblockN{Jan Skoglund}
\IEEEauthorblockA{\textit{Google LLC}\\
San Francisco, CA \\
jks@google.com}

}

\maketitle

\begin{abstract}
In the history of audio and acoustic signal processing, perceptual audio coding has certainly excelled as a bright success story by its ubiquitous deployment in virtually all digital media devices, such as computers, tablets, mobile phones, set-top-boxes, and digital radios. From a technology perspective, perceptual audio coding has undergone tremendous development from the first very basic perceptually driven coders (including the popular mp3 format) to today’s full-blown integrated coding/rendering systems. This paper provides a historical overview of this research journey by pinpointing the pivotal development steps in the evolution of perceptual audio coding.  Finally, it provides thoughts about future directions in this area.
\end{abstract}



\section{{Introduction}}

Looking back at the history of audio and acoustic signal processing technology, audio coding serves as a successful exemplar, accomplishing a virtuous cycle through collaboration between academic and industrial researchers, their standardization efforts, and their application to commercial products. Billions of people use audio coding technology to compress audio signals daily, from digital broadcasts and streaming/storing digital music to communication in diverse acoustic environments.  To this end, researchers have aimed at providing ``good" to even ``transparent" sound quality to human listeners in various computing environments (e.g., PCs and smartphones), playback configurations (e.g., earbuds and loudspeakers), network setups (e.g., different wireless network protocols), and finally, various listeners' subjectivity in defining sound quality.

One of the key principles that led to the technological and commercial success of audio coding is the notion of \textit{perceptual} audio coding, which aims to preserve the coded signal's quality to the degree that it is difficult for the human auditory system to discern the difference between before and after the compression. This guiding principle has encouraged researchers to develop numerous algorithms that have successfully reduced the number of bits to represent the audio signal without significantly degrading the perceived sound quality. 

In this paper, we provide a condensed review of the history of perceptual audio coding. Sec. \ref{sec:basic} covers a few basic concepts that distinguish perceptual codecs from their counterparts. Sec. \ref{sec:history} follows to present a chronological overview of important technological advancements in the history of audio coding, such as the initial transform coders with perceptual bit-allocation (actually ``noise allocation") algorithms, e.g., the ``mp3" codec, perceptual multi-channel coding, low-delay variants, spectral bandwidth extension, parametric coding for multi-channel and multi-object signals, and unified codecs that handle both speech and audio signals. Sec. \ref{sec:history} also presents more advanced coding scenarios, such as spatial coding and its use in AR/VR applications. Finally, Sec. \ref{sec:summary} summarizes the paper and also discusses one of the future directions of audio coding, that of incorporating data-driven approaches and machine learning, and the challenges incumbent on that approach.

\begin{figure}
     \centering
     \begin{subfigure}[b]{0.46\columnwidth}
         \centering
         \vspace{-0.1in}
         \includegraphics[width=\textwidth]{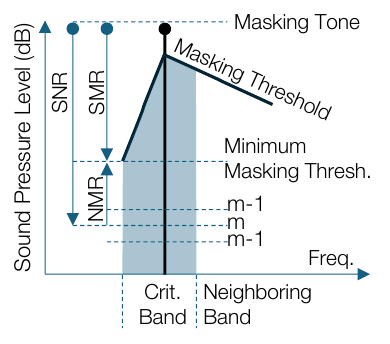}\vspace{-0.1in}
         \caption{}
         \label{fig:simul_masking}
     \end{subfigure}
     \hfill
     \begin{subfigure}[b]{0.51\columnwidth}
         \centering
         \vspace{-0.1in}
         \includegraphics[width=\textwidth]{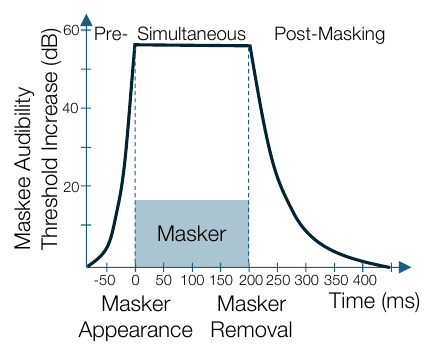}\vspace{-0.1in}
         \caption{}
         \label{fig:temp_masking}
     \end{subfigure}
        \caption{(a) The simultaneous masking phenomenon (b) temporal masking. Figures were adapted from \cite{PreEcho}.}
        \label{fig:masking}
\end{figure}

\section{Basic Concepts}\label{sec:basic} 

\subsection{{Psychoacoustics and Masking}}\label{sec:psycho}

Psychoacoustics is an interdisciplinary scientific area that studies the human auditory system's perception and interpretation of different sounds, such as the perception of pitch from physical frequencies, loudness that depends on amplitude and frequency, localization of spatial sound, and masking between different sound components. 

Masking describes the phenomenon that louder sounds hide, i.e., ``mask", softer sounds in their spectral or temporal vicinity.  Fig. \ref{fig:simul_masking} illustrates how the threshold of masking has a shallow slope towards higher and a steeper slope towards lower frequencies, often expressed on a ``critical band" scale. Tonal signals are weaker maskers than noise-like signals. Similarly, Fig. \ref{fig:temp_masking} depicts temporal masking before the onset and after the end of a switched masker signal.
Exploiting these masking phenomena is of central importance in perceptual audio coding to hide the coding error signal.

\subsection{{Waveform Coding} }
In a straightforward digital representation of the audio waveform, each signal sample is quantized individually and represented by a digital code word, a technique traditionally called pulse-code modulation (PCM) coding. However, since there is a high sample-level correlation in most audio signals, higher compression efficiency can be obtained by so-called predictive coding. In differential PCM (DPCM), which was first patented in the 1950s, a prediction of the next incoming sample is first made based on previous quantized samples. The difference between the prediction and the incoming sample is then quantized and transmitted, resulting in a lower quantization error variance. Often this prediction is just the previous quantized sample, but it can also be formed as a linear combination of several previously quantized samples. Later on, adaptive DPCM (ADPCM) \cite{JayantNoll} was shown to provide even higher coding efficiency by adapting both the prediction and quantization to the signal.
Another way of utilizing this correlation is through transform coding. In this technique, a (typically linear) transform is applied to blocks of successive time samples. The resulting transform coefficients are then individually quantized.

\begin{figure}
    \centering
    \includegraphics[width=.9\linewidth]{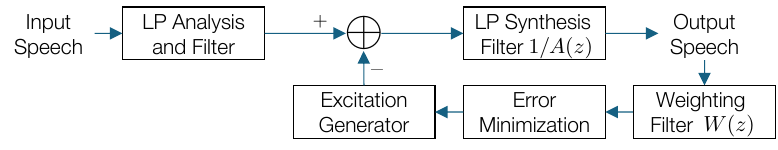}
    \vspace{-0.05in}
    \caption{Linear predictive analysis-by-synthesis coding}
    \vspace{-0.1in}
    \label{fig:LPAS}
\end{figure}

\subsection{{Speech Coding}}
Before its use in entertainment, audio coding's major application was for speech communication through low bandwidth channels such as telephony and secure radio communication.
Speech signals generally have a band-limited frequency content, i.e., the long-term average spectrum decays rapidly after 4-5 kHz. Sampling rates of 8-10 kHz, therefore, became the natural choice for digitizing speech in telephony applications. For example, the ITU\mbox{-}T
G.711 standard in public switched telephone networks has a sampling frequency of 8 kHz using PCM coding. Note that a narrow bandwidth of 4 kHz is generally sufficient for intelligible conversations, but for natural speech fidelity, higher sampling frequencies are necessary. 

Early on, speech codecs were designed to match waveforms as much as possible by minimizing the Euclidean distance between the incoming samples and the quantized ones, i.e., maximizing signal-to-noise ratio (SNR). This leads to a quantization noise that has a flat spectrum. Simultaneous masking allows for achieving higher perceptual quality, e.g., by shaping the quantization noise to approximately follow the signal spectrum \cite{Weighting}.
Linear predictive analysis-by-synthesis (LPAS) coders, such as the first GSM coder \cite{RPE} and the groundbreaking CELP coder \cite{CELP}, utilized such a weighted error criterion. In LPAS coding, depicted in Fig.~\ref{fig:LPAS}, the time-domain speech signal, $s(n)$, is modeled as an excitation signal, $u(n)$, feeding a linear predictive synthesis filter, $1/A(z)$.
The encoding criterion is to find the best excitation parameters such that the variance of the filtered difference signal between the original and coded signal is minimized. The weighting filter, $W(z)$, is 
often based on the linear prediction filter as $ W(z) =  \frac{A(z/\rho_1)}{A(z/\rho_2)} $, where $0<\rho_2<\rho_1<1$. 

Today, speech coding is no longer restricted to narrowband telephony. Higher signal bandwidths have applications such as video conferencing over the Internet. This has paved the way for wideband speech coding, using 16 kHz sampling frequency. LPAS coding has also been shown to be successful for wideband coding and is found in most of the recent speech coding standards.

\subsection{{Perceptual Audio Coding}}\label{sec:perceptual_basic}

As implied by the name, perceptual audio coding focuses on optimizing the subjective (perceived) sound quality rather than a mathematical error metric, such as 
SNR. The main strategy is to shape the coding error (caused by a bitrate-limited resolution) in time and frequency such that it is below the masking threshold, as explained in Sec. \ref{sec:psycho}. In this way, the error is rendered inaudible, or it is perceived as minimally objectionable if the overall bitrate is too low to achieve this goal. This approach is called \textit{noise shaping} and intentionally accepts some decrease in SNR as a side effect, as SNR is not what the human auditory system reacts to.

\begin{figure}
    \centering
    \includegraphics[width=.9\columnwidth]{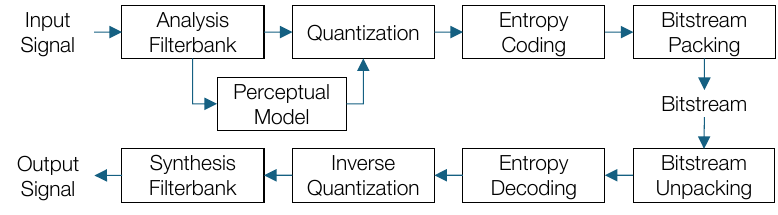}
    \vspace{-0.05in}
    \caption{Perceptual audio encoder (top) and decoder (bottom).}
    \vspace{-0.1in}
    \label{fig:perceptual_coder}
\end{figure}

In contrast to time-domain ADPCM-style waveform coding, perceptual audio coding is typically based on a filterbank-based architecture \cite{Krasner} as explained in Fig. \ref{fig:perceptual_coder}. In the encoder, the input time domain signal is converted into subsampled spectral coefficients by an analysis filterbank in a framewise fashion where transforms are understood as specific perfectly reconstructing filterbanks. The spectral coefficients are then quantized and coded and put as binary fields into a bitstream multiplex (e.g., an ``.mp3'' file). What makes it a perceptual audio coder is that a perceptual model analyzes each frame of the input signal to determine its time- and frequency-dependent masking threshold. This estimate is used to drive the quantization process so that the masking requirements are met as well as possible (i.e., the quantization error is below the masking threshold) while the overall bitrate requirements are also met. This is a constrained optimization problem for which several approaches have been investigated \cite{QCtutorial}.

The decoder's structure mirrors the encoder's without a perceptual model. First, the bitstream is disassembled, and then the coded spectral coefficients are decoded and scaled back to their original value range. Finally, the synthesis filterbank maps the reconstructed spectral coefficients back into a time domain waveform for playback.


\section{Historical Timeline}\label{sec:history}

\subsection{{The First Generation of Perceptual Audio Coders}}

The first generation of perceptual audio coders was built according to the simple structure described in Sec. \ref{sec:perceptual_basic}, i.e., with an analysis filterbank, perceptual model, quantization/coding, and bitstream multiplexer as the building blocks. A particular challenge was the choice of the optimal filterbank, which determines the inherent time/frequency resolution of the coder and, thus, its coding performance on transient and tonal signals. The former requires a high time resolution to avoid ``pre-echo" \cite{PreEcho} artifacts, while the latter necessitates a high frequency (and thus low time) resolution.

Therefore, the first-generation coders can be roughly divided into coders with a high frequency resolution (e.g., 1024 or more bands) and those with a low frequency resolution (e.g., up to 256 spectral bands) based on their filterbank configuration. As for the high frequency resolution coders, a modified discrete cosine transform (MDCT) \cite{MDCT} filterbank with window switching \cite{SWITCH} turned out to be the best performing and most popular solution, whereas coders with lower frequency resolution often preferred polyphase filterbanks.

The perceptual models of the coders varied in the degree of accuracy, in which the calculation of the masking threshold was performed (e.g., the calculation of the signal’s tonality). The quantization of the spectral coefficients was carried out individually (e.g., scalar quantization) with either uniform or non-uniform bins. Solutions for coding of the quantized coefficients ranged from simple block companding \cite{JayantNoll} to four-dimensional Huffman coding.

Additionally, some computationally simple approaches were developed for joint coding of the spectral coefficients of stereophonic signals. The most popular approaches are the mid/side (M/S) stereo \cite{MS} and intensity stereo \cite{Intens}, exploiting redundancy between the channels while avoiding binaural unmasking problems \cite{MS2} \cite{BMLD}.

Some important milestones in this phase were, e.g., ASPEC \cite{ASPEC}, MPEG-1 Layers 1, 2, and 3 \cite{MPEG1} and AC-2 \cite{AC-23}, where MPEG-1 Layer~3 found widespread applications in music download, streaming and playback on mobile devices. Under its short name ``mp3”, it became the most well-known compressed format worldwide with an estimated over a trillion existing bitstreams and many billions of devices that can play back the format.
This first coder generation set the stage for all later developments, enabling the coding of mono or stereo signals. It typically provided transparent/good audio quality for stereo signals in the range of 256/192 kbps (MPEG-1 Layer 2) down to 192/128 kbps (MPEG-1 Layer 3, a.k.a. mp3).

\subsection{{Perceptual Multi-Channel Coding}}

The next major step after mono and stereophonic coding was to enable efficient coding of multi-channel sounds, such as sound produced for 5.1 loudspeaker setups. This required a deepened understanding of both the available redundancy between the channel signals and the perceptual requirements for transparent coding (i.e., spatial masking, including effects such as binaural masking level difference \cite{BMLD}). While initial stereo-compatible coding methods required very high bitrates due to masking issues, MPEG-2 advanced audio coding \cite{AAC} is an example of a successful multi-channel coder, requiring only half the rate of the stereo-compatible codec for achieving broadcast quality and significantly less rate than AC-3 \cite{DD,CRC}. Its good performance results from a better variable time/frequency resolution, bandwise M/S stereo coding, enhanced entropy coding for spectral coefficients and side information, and a range of specific coding tools.

\subsection{{Beyond Bitrate and Channels}}

At this point of development, it became clear that merely enhancing the coder’s efficiency (i.e., requiring even less bitrate) was insufficient to bring perceptual audio coding into all conceivable application domains. For example, using audio coding for wireless transmission called for enhanced error robustness of the bitstreams \cite{MP4Audio}. Broadcast streaming over variable rate channels like the Internet suggested codecs' ability to adapt the bitrate in real-time to the instantaneous transmission capacity or using scalable coding techniques \cite{Scalab,MP4Audio}. Furthermore, using codecs like mp3 or AAC for high-quality telecommunication is not possible due to their inherent delay of several hundred milliseconds. Thus, low-delay perceptual audio coding was introduced, leading, e.g., to low-delay derivatives of AAC \cite{LD-AAC,MP4Audio}. At this point, purely parametric coding approaches were also tried out for their merits at very low bitrates, modeling sound with sinusoids and noise and allowing pitch and speed change during decoding \cite{HILN,MP4Audio}. After some more focused development of purely parametric coding techniques \cite{ARDOR,SSC}, it turned out that filterbank-based audio coding remained superior due to its attractive trade-off in bitrate and computational efficiency.

\subsection{{Bandwidth Extension}}

Meanwhile, the spread of cellular networks and the push for digital broadcasting encouraged perceptual audio coding to lower bitrates even further (below 64 kbps per stereo signal). However, coding enough spectral coefficients with appropriate resolution created a barrier to decreasing the bitrate when high audio quality was required. The only way of achieving very low bitrates was to sacrifice the quality and--most importantly--bandwidth of the transmitted audio. Bandwidth extension (BWE) techniques overcame this limitation by transmitting a low-bandwidth base audio signal and reconstructing the missing high-frequency range from the base signal and very compact parametric side information (e.g., 1-2 kbps per channel) that allows approximating the original time/frequency and perceptual characteristics. Since both a transmitted waveform and parametric side information are used, this can be considered a hybrid of waveform and parametric coding rather than pure parametric coding. While these techniques do not provide fully transparent coding, they still offer quite a good sound at very low rates (e.g., 48 kbps stereo) and can be considered a major breakthrough in modern audio coding. The most well-known BWE scheme is spectral band replication (SBR) \cite{SBR}, which was standardized in 2003 and excelled through its very efficient processing structure. It was combined with the family of AAC coders and became very successful in the market \cite{IEEE HE-AAC}.

\subsection{{Parametric Coding of Several Channels}}\label{sec:pcc}

Similar to the efficiency break-through by hybrid waveform/parametric BWE, another milestone was achieved soon afterward: while previous joint coding efforts for several channel signals had shown either a very limited, signal-dependent gain (M/S stereo coding) or led to strong artifacts for critical signals (intensity stereo coding), a new generation of techniques focused on the systematic exploitation of frequency-dependent spatial listening cues \cite{BMLD}, including inter-aural level differences (IALDs), inter-aural time/phase differences (IATDs/IAPDs) and inter-aural cross-correlation (IACC). The central idea was to reduce several channels to a common (mono or stereo) downmix signal, which is transmitted by regular perceptual coding methods. Additionally, the spatial cues of the original signal are extracted into very compact parametric side information and used in the decoder to guide the upmix process that recovers the original number of channels from the downmix signal while the original spatial cues are approximately preserved. This ``spatial audio coding" approach was successfully developed for 2-1-2 (stereo-to-mono-to-stereo) processing (Parametric Stereo \cite{PS}) and multi-channel operation (5-1-5, 5-2-5 etc., \cite{Faller/Breebart Book}  \cite{MPS}, \cite{JCC}). Examples of audio coders using this approach include \cite{IEEE HE-AAC} \cite{DDP} \cite{USAC}. Similarly to BWE, this hybrid between waveform coding of the downmix and parametric upmixing brought marked improvements for very low bitrates (e.g., 32~kbps stereo or 5.1 sound at 64 kbps), although full transparency cannot be expected. 


\subsection{{Parametric Coding of Objects}}

Representing complex audiovisual content by ``objects" (sources) and a description of how to combine them (``scene description") is a far-reaching concept that was proposed already before 2000 but was not economically viable at that time (MPEG-4 \cite{MPEG-4}). It encodes content independently of its eventual consumption environment (e.g. number of loudspeakers) and allows for real-time interactive rendering.

Audio objects can be parametrically coded, building on concepts of parametric coding of multiple channels described in Sec. \ref{sec:pcc}. Standardized in 2010, MPEG-D spatial audio object coding \cite{SAOC} allows bitrate-efficient representation and interactive rendering of object-based content. Like in parametric coding of multiple channels, the input (here: object) signals are converted into a mono or stereo downmix plus compact parametric side information that represents the characteristics of the input on a time/frequency grid. For objects, this analogy of the ``multi-channel cue side information" is the relations of the object levels and their mutual correlation values. Applications include an interactive remix of music content (e.g., producing a Karaoke remix for sing-along) and adjustable rendering of teleconferences for optimum talker balance, spatial placement, and thus highest intelligibility. Similar principles are implemented in \cite{JOCC}.


\subsection{{Combined Audio/Speech Coding }}

The source coding methods in speech coders typically provide much lower bitrates than the audio coders do. A number of coders have combined these two paradigms to use speech coder tools when the input signal is ``speech-like" and audio coder tools otherwise. 

Extended adaptive multi-rate--wideband (AMR-WB+) \cite{TS 26.290} is an extension of AMR-WB \cite{{AMR-WB}}. It adds support for stereo signals and higher sampling rates. It incorporates algebraic code excited linear prediction (ACELP) speech tools and transform coded excitation (TCX) audio tools. Automatic switching between TCX and ACELP  provides good quality for all types of signals at moderate bit rates.  AMR-WB+ is backward compatible with AMR-WB.

MPEG unified speech and audio coding (USAC) \cite{USAC} builds on the AAC architecture and adds TCX tools for speech-like signals and ACELP for very speech-like signals. It deploys content-driven on-the-fly switching between any of the three coding architectures.

Enhanced voice services (EVS) \cite{EVS} operates in a broad range of bitrates, is highly robust against packet loss, and is appropriate for low-delay real-time communication systems. Like USAC, it can switch between speech and audio compression to obtain seamless and reliable operation with a low algorithmic delay of 32 ms. 

The open-source Opus \cite{Opus} is a versatile coder supporting sampling rates from 8~kHz to 48~kHz, 1 to 255 channels, and bitrates from 8~kbps to 510~kbps, at frame sizes from 5~ms to 60~ms. It is based on a combination of the linear predictive coding
part (SILK \cite{SILK}) and the transform coding part (CELT \cite{CELT}).

\subsection{{Generic Spatial Coding}}

For a more immersive experience, MPEG-H 3D Audio, standardized as ISO/IEC 23008-3, supports the coding of many audio objects, channel signals, and higher order ambisonics (HOA) signals  \cite{3DA}. Considering that channel signals are objects at a specific location, the coding of audio objects builds upon previous stereo-coding technology. However, the very large number of components of HOA signals required an innovative process in which the HOA sound field is decomposed into predominant and ambient sound components. At the same time, parametric side information is generated to enable the decoder to reconstruct the original HOA signal. 
MPEG-H 3D Audio supports rich metadata that allows for a very flexible presentation of the audio program, including personalized presentations, advanced loudness management, interactive dialog enhancement, and an engine to binauralize the audio program for presentation on headphones. 
Dolby AC-4 \cite{AC4}, standardized as ETSI TS 103 190, supports object coding and enables video-frame-synchronous coding. Opus also has a mode for coding HOA \cite{OpusHOA}.

\subsection{{Coding/Rendering for AR/VR}}

Recently, the coding and rendering of audio for virtual, augmented, and mixed reality (AR/VR/MR) has gained great interest. The MPEG\mbox{-}I immersive audio standard (ISO 23090-4) specifies the coding and rendering of audio for virtual acoustic scenes with six degrees of freedom (6DoF) for the user \cite{MPEG-I}. It can compactly represent the required information to enable efficient distribution over severely bitrate-limited channels. In the system, audio waveform compression is provided by the MPEG-H 3D Audio codec, providing efficient use of audio objects, channel signals, and HOA material in virtual audio scenes. Additionally, the system receives metadata about the virtual environment (``scene"), such as acoustic properties of the room (e.g., reflection, absorption, etc.) and room geometry (e.g., doors and possible sound-occluding objects). The immersive audio renderer receives audio signals from the 3D audio decoder and continuously tracks the position and orientation of the user (6DoF) from a local input. The rendering process includes detailed modeling of room acoustics and complex acoustic phenomena, e.g., occlusion and diffraction due to acoustic obstacles and Doppler effects, as well as real-time interactivity with the user, e.g., opening and closing doors or manipulating audio sources.

\begin{figure}
    \centering
    \includegraphics[width=0.95\linewidth]{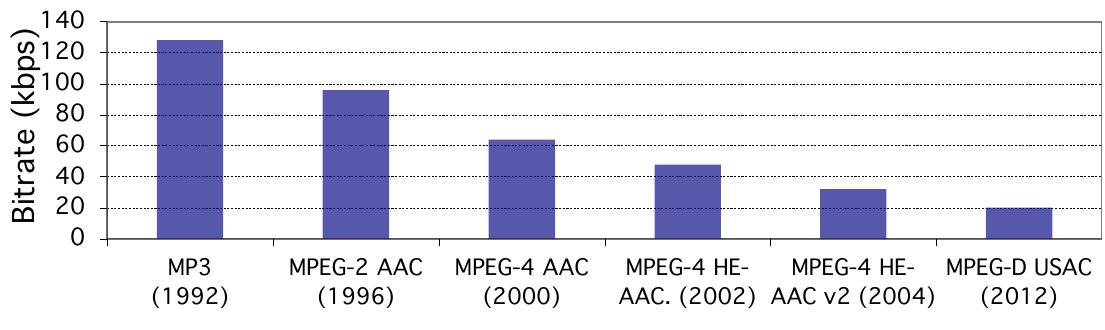}\vspace{-0.1in}
    \caption{Progress over time: bitrates required for ``good quality" (not transparent) coding of stereo audio.}
    \label{fig:bitrate_graph}
\end{figure}

\section{{Summary, Future Perspectives and Challenges}}\label{sec:summary}

During the past four decades, perceptual audio coding has undergone a tremendous evolution, both in terms of technological development and practical deployment. While the first generation of audio coders required 128-256 kbps for representing a stereo audio signal at good quality, current state-of-the-art codecs provide that quality at 20 kbits/s and below (see Fig. \ref{fig:bitrate_graph}), and some of them unite the worlds of perceptual audio and source-oriented speech coding. Beyond mere coding efficiency, today’s codecs provide many more functionalities like low-delay modes for high-quality communication, error resilience for transmission over error-prone (e.g., wireless) channels, as well as coding and rendering of surround and 3D audio content with up to 22.2 channels, many objects, and HOA. Most recently, audio coding technology is being combined with binaural rendering for VR/AR content.

Application-wise, perceptual audio coding has truly revolutionized the way we distribute, buy, and consume digital media. Starting with mp3, music has been freed from being tied to a physical medium, such as LP records or CDs. Billions of media devices exist that play compressed bitstreams in mp3 format, or one of its successors, and trillions of existing bitstreams illustrate the ubiquity of music content. Despite many predictions that audio coding would soon become irrelevant due to the enormous resources in storage capacity and network speed available today, audio coding has continued to thrive and progress to accommodate the global trend toward wireless services and an ever-increasing number of service programs offered.

Although the future of perceptual audio coding can be multifaceted, data-driven machine-learning approaches to coding are increasingly popular and noticeable. A neural audio coding system can be formulated as an autoencoder. Hence, while it learns to recover the input signal in an end-to-end fashion, it also learns the low-dimensional and quantized ``code" in its bottleneck \cite{Kankanahali, SoundStream}. Meanwhile, a neural decoder can employ a generative model to ``synthesize" waveforms from ultra-low bitrate ($<$ 2 kbps) codes \cite{WaveNet, VQ-VAE}. Although these neural codecs have pushed the limits of audio coding by showing greater coding gain, they tend to increase the complexity, making their use in resource-constrained devices expensive. Hence, some neural codecs incorporate conventional coding methods to improve their computational efficiency, e.g., a merger of recurrent neural networks and LPC \cite{LPCNet} and psychoacoustically modified loss functions for transparent audio coding \cite{Kai}, etc. Another challenge is that most successful neural codecs are specialized in speech, while neural audio compression needs more generalization effort. Finally, an increasingly popular usage of neural codecs is as input to other neural network-based downstream tasks, such as speech synthesis \cite{VALLE}, speech recognition \cite{HuBERT}, speech enhancement \cite{MIIPHER}, etc. A recent review of neural speech and audio coding is available in \cite{NSAC}.

In summary, it can be said that perceptual audio coding has truly revolutionized many parts of our modern life and, most noticeably, made music available seamlessly in any location at any time.


\begin{thebibliography}{00}
\bibitem{JayantNoll} N. Jayant and P. Noll, ``Digital Coding of Waveforms: Principles and Applications to Speech and Video", Prentice-Hall, 1984.

\bibitem{Weighting} B. Atal and M. Schroeder, ``Predictive Coding of Speech Signals and Subjective Error Criteria", IEEE Transactions on Acoustics, Speech, and Signal Proc., Vol. 27, No. 3, pp. 247-254, 1979. 

\bibitem{PreEcho}  T. Painter and A. Spanias, ``Perceptual Coding of Digital Audio,” Proceedings of IEEE, Vol. 88, No. 4, pp. 451-513, April 2000.

\bibitem{RPE} P. Kroon, E. Deprettere, and R. J. Sluyeter, ``Regular-Pulse
Excitation--A Novel Approach to Effective and Efficient Multipulse Coding of Speech,” IEEE Trans. on  Acoustics, Speech, and Signal Proc., vol. ASSP-34, no. 5, 1986

\bibitem{CELP} M. Schroeder and B. Atal, ``Code-Excited Linear Prediction (CELP): High-Quality Speech at Very Low Bit Rates," IEEE Int'l Conf. on Acoustics, Speech, and Signal Proc., 1985.

\bibitem{Krasner}  M. A. Krasner, ``Digital Encoding of Speech and Audio Signals based on the Perceptual Requirements of the Auditory System", Tech. Report 535, MIT Lincoln Lab., Lexington 1979.

\bibitem{QCtutorial} J. Herre, ``Temporal Noise Shaping, Quantization and Coding Methods in Perceptual Audio Coding---A Tutorial Introduction", the 17th Int'l Audio Eng. Soc. Conf. on High Quality Audio Coding, 1999. 



\bibitem{MDCT} J. Princen, A. Johnson and A. Bradley, ``Subband/Transform Coding Using Filter Bank Designs Based on Time Domain Aliasing Cancellation", IEEE Int'l Conf. on Acoustics, Speech, and Signal Proc., pp. 2161-2164, 1987.
\bibitem{SWITCH}  B. Edler, ``Codierung von Audiosignalen mit überlappender Transformation und adaptiven Fensterfunktionen", Frequenz, Vol. 43, pp. 252-256, 1989 (in German).

\bibitem{MS}  J. D. Johnston and A. J. Ferreira, ``Sum-Difference Stereo Transform Coding", IEEE Int'l Conf. on Acoustics, Speech, and Signal Proc., pp. 569-571, 1992.
\bibitem{Intens}  J. Herre, K. Brandenburg and D. Lederer, ``Intensity Stereo Coding", the 96th Audio Eng. Soc. Convention, 1994, Preprint 3799. 
\bibitem{MS2}  J. Herre, K. Brandenburg, and E. Eberlein, ``Combined Stereo Coding", the 93rd Audio Eng. Soc. Convention, 1992, Preprint 3369. 

\bibitem{BMLD} J. Blauert, ``Spatial Hearing: the Psychophysics of Human Sound Localization (revised edition)", MIT Press, 1997.

\bibitem{ASPEC}  K. Brandenburg et al., ``ASPEC: Adaptive Spectral Perceptual Entropy Coding of High Quality Music Signals", the 90th. Audio Eng. Soc. Convention, 1991, Preprint 3011. 
\bibitem{MPEG1}  K. Brandenburg et al., ``The ISO/MPEG-Audio Codec: A Generic Standard for Coding of High Quality Digital Audio", the 92nd Audio Eng. Soc. Convention, 1992, Preprint 3336. 
\bibitem{AC-23}  L. Fielder et al., ``AC-2 and AC-3: Low-Complexity Transform-Based Audio Coding", in Collected Papers on Digital Audio Bit-Rate Reduction, N. Gilchrist and C. Grewin, Eds. (Audio Eng. Soc., New York, 1996), pp. 54-72. 

\bibitem{AAC} M. Bosi et al., ``ISO/IEC MPEG-2 Advanced Audio Coding", J. of Audio Eng. Soc., Vol. 45, No. 10, 1997, pp. 789-814. 
\bibitem{DD} M. Davis, ``The AC-3 Multichannel Coder", the 95th Audio Eng. Soc. Convention, 1993, Preprint 3774. 
\bibitem{CRC} G. Soulodre et al., ``Subjective Evaluation of State-of-the-Art Two-Channel Audio Codecs", J. of Audio Eng. Soc., Vol. 46, No. 3, pp. 164-177, 1998. 
\bibitem{MP4Audio}  J. Herre, H. Purnhagen, ``General Audio Coding", in F. Pereira, T. Ebrahimi (Eds.), "The MPEG-4 Book", pp. 487-544, Prentice Hall IMSC Multimedia Series, 2002; ISBN 0-13-061621-4
\bibitem{Scalab} B. Grill, ``A Bit Rate Scalable Perceptual Coder for MPEG-4 Audio," the 103rd Audio Eng. Soc. Convention, 1997, Preprint 4620. 
\bibitem{LD-AAC} E. Allamanche et al., ``MPEG-4 Low Delay Audio Coding based on the AAC Codec", the 106th Audio Eng. Soc. Convention, 1999, Preprint 4929. 
\bibitem{HILN}  H. Purnhagen, and N. Meine, ``HILN -- The MPEG-4 Parametric Audio Coding Tools," IEEE Int'l Symposium on Circuits And Systems (ISCAS), 2000. 
\bibitem{ARDOR}  N. van Schijndel et al., ``Adaptive RD Optimized Hybrid Sound Coding", J. of Audio Eng. Soc., Vol. 56, No. 10, pp. 787-809, 2008. 
\bibitem{SSC}  E. Schuijers et al., ``Advances in Parametric Coding for High-Quality Audio", the 114th Audio Eng. Soc. Convention, 2003. 
\bibitem{SBR}  M. Dietz et al., ``Spectral Band Replication---A Novel Approach in Audio Coding", the 112th Audio Eng. Soc. Convention, 2002, Preprint 5553. 
\bibitem{IEEE HE-AAC}  J. Herre and M. Dietz, ``Standards in a Nutshell: MPEG-4 High-Efficiency AAC Coding", IEEE Signal Proc. Magazine, Vol. 25, Issue 3, pp 137 - 142, 2008. 

\bibitem{PS} E. Schuijers et al.,  ``Low-Complexity Parametric Stereo Coding", the 116th Audio Eng. Soc. Convention, 2004, Preprint 6073. 

\bibitem{Faller/Breebart Book}  J. Breebaart and C. Faller, ``Spatial Audio Processing: MPEG Surround and Other Applications", John Wiley \& Sons, Nov. 2008.
\bibitem{MPS}  J. Herre et al., ``MPEG Surround – The ISO/MPEG Standard for Efficient and Compatible Multichannel Audio Coding", J. of Audio Eng. Soc., Vol. 56, No. 11, 2008, pp. 932-955. 
\bibitem{JCC} H.-M. Lehtonen et al., ``Parametric Joint Channel Coding of Immersive Audio”, the 142th Audio Eng. Soc. Convention, Paper No. 9740, 2017. 

\bibitem{DDP} L. D. Fielder et al., ``Introduction to Dolby Digital Plus, an Enhancement to the Dolby Digital Coding System," the 117th Audio Eng. Soc. Convention, 2004. Preprint 6196. 

\bibitem{USAC} M. Neuendorf et al.: ``The ISO/MPEG Unified Speech and Audio Coding Standard---Consistent High Quality for All Content Types and at All Bit Rates", J. of Audio Eng. Soc., Vol. 61, No. 12, 2013. 
\bibitem{MPEG-4} F. Pereira, T. Ebrahimi (Eds.), "The MPEG-4 Book", pp. 487-544, Prentice Hall IMSC Multimedia Series, 2002; ISBN 0-13-061621-4.

\bibitem{SAOC} J. Herre et al., ``MPEG Spatial Audio Object Coding---The ISO/MPEG Standard for Efficient Coding of Interactive Audio Scenes", J. of Audio Eng. Soc., Vol. 60, No. 9, 2012, pp. 655-673. 


\bibitem{JOCC} H. Purnhagen et al., ``Immersive Audio Delivery Using Joint Object Coding”, the 140th Audio Eng. Soc. Convention, Paper No. 9587, 2016. 

\bibitem{TS 26.290} 3GPP TS 26.290; Audio Codec Processing Functions; Extended Adaptive Multi-Rate--Wideband (AMR-WB+) Codec; Transcoding Functions.

\bibitem{AMR-WB} B. Bessette et al., ``The adaptive multirate wideband speech codec (AMR-WB)", IEEE Trans. on Speech and Audio Proc., vol. 10, no. 8, pp. 620-636, 2002.

\bibitem{EVS} M. Dietz et al., ``Overview of the EVS Codec Architecture",  IEEE Int'l Conf. on Acoustics, Speech, and Signal Proc., pp. 5698-5702, 2015.

\bibitem{Opus} J.-M. Valin, K. Vos, and T. B. Terriberry, “Definition
of the Opus Audio Codec,” IETF RFC 6716, 2012.

\bibitem{SILK} K. Vos, K. V. Sørensen, S. S. Jensen, and J.-M. Valin, “Voice coding with Opus,” the 135th Convention of the Audio Eng. Soc., 2013.

\bibitem{CELT} J.-M. Valin et al., ``High quality, low-delay music coding in the Opus codec," the 135th Audio Eng. Soc. Convention, 2013. 




\bibitem{3DA} J. Herre, J. Hilpert, A. Kuntz, and J. Plogsties, ``MPEG-H Audio—The New Standard for Universal Spatial/3D Audio Coding," J. of Audio Eng. Soc., Vol. 62, No. 12, 2014, pp. 821–830. 



\bibitem{AC4} K. Kjörling et al., ``AC-4---The Next Generation Audio Codec", the 140th Audio Eng. Soc. Convention, 2016. Preprint 9491. 

\bibitem{OpusHOA} J. Skoglund, M. Graczyk, ``Ambisonics in an Ogg Opus container", IETF RFC 8486, 2018.


\bibitem{MPEG-I} J. Herre and S. Disch, ``MPEG-I Immersive Audio -- Reference Model for the Virtual/Augmented Reality Audio Standard," J. of Audio Eng. Soc., Vol 71, No. 5, pp. 229-240, 2023. 








\bibitem{Kankanahali} S. Kankanahalli, ``End-to-end Optimized Speech Coding with Deep Neural Networks,” in Proc. of IEEE Int'l Conf. on Acoustics, Speech, and Signal Proc., 2018.

\bibitem{SoundStream} N. Zeghidour et al., ``Soundstream: an End-to-End Neural Audio Codec,” IEEE/ACM Trans. Audio, Speech and Lang. Proc., vol. 30, p. 495–507, 2022.

\bibitem{WaveNet} W.B.Kleijn et al., ``WaveNet Based Low Rate Speech Coding," in Proc. of IEEE Int'l Conf. on Acoustics, Speech, and Signal Proc., 2018.

\bibitem{VQ-VAE} C. Garbacea et al.,  ``Low Bit-Rate Speech Coding with VQ-VAE and a WaveNet Decoder,” in Proc. of IEEE Int'l Conf. on Acoustics, Speech, and Signal Processing, 2019.

\bibitem{LPCNet} J.-M. Valin and J. Skoglund, ``A Real-Time Wideband Neural Vocoder at 1.6kb/s Using LPCNet,” Interspeech, 2019.

\bibitem{Kai} K. Zhen et al., ``Psychoacoustic Calibration of Loss Functions for Efficient End-to-End Neural Audio Coding," IEEE Signal Proc. Letters, vol. 27, pp. 2159–2163, 2020.


\bibitem{VALLE} C. Wang et al., ``Neural Codec Language Models are Zero-Shot Text to Speech Synthesizers," arXiv:2301.02111.

\bibitem{HuBERT} A. Polyak et al., ``Speech Resynthesis from Discrete Disentangled Self-Supervised Representations,” Interspeech, 2021.

\bibitem{MIIPHER} Y. Koizumi et al., ``Miipher: A Robust Speech Restoration Model Integrating Self-Supervised Speech and Text Representations," IEEE Workshop on Applications of Signal Proc. to Audio and Acoustics, 2023.

\bibitem{NSAC} M. Kim and J. Skoglund, ``Neural Speech and Audio Coding," IEEE Signal Processing Magazine, 2024 (to appear).


\end{thebibliography}
\end{document}